\documentclass[manuscript]{acmart}
\AtBeginDocument{%
  \providecommand\BibTeX{{%
    \normalfont B\kern-0.5em{\scshape i\kern-0.25em b}\kern-0.8em\TeX}}}

\setcopyright{acmcopyright}
\copyrightyear{2025}
\acmYear{2025}
\acmDOI{10.1145/3706598.3713838}

\acmConference[CHI '25]{ACM CHI Conference on Human Factors in Computing
Systems}{April 26 - May 1, 2025}{Yokohama, Japan}
\acmBooktitle{Proceedings of the CHI Conference on Human Factors in Computing Systems (CHI '25), April 26 - May 1,
 2025, Yokohama, Japan}

\usepackage{graphicx}
\usepackage{caption}
\usepackage{subcaption}
\usepackage{booktabs}
\usepackage[normalem]{ulem}
\useunder{\uline}{\ul}{}

\newcommand{\beginsupplement}{
    \setcounter{table}{0}
    \renewcommand{\thetable}{S\arabic{table}}
    \setcounter{figure}{0}
    \renewcommand{\thefigure}{S\arabic{figure}}
}

\newcommand{\codesignparticipants}{
\begin{table}[]
\centering
\def\arraystretch{1.5}
\begin{tabular}{@{}|l|l|l|l|l|@{}}
\hline
\textbf{ID} & \textbf{Gender} & \textbf{Race and Ethnicity} & \textbf{Age} & \textbf{Job Level} \\ \hline
D1 & Male & Asian & 40-44 & First-level management \\ \hline
D2 & Female & White (non-Hispanic) & 20-24 & Intermediate level \\ \hline
D3 & Male & Asian & 20-24 & Entry level \\ \hline
D4 & Female & Asian (Korean) & 25-29 & First-level management \\ \hline
D5 & Male & Asian & 20-24 & Entry level \\ \hline
D6 & Female & White & 25-29 & Intermediate level \\ \hline
D7 & Male & Asian (Indian) & 25-29 & Intermediate level \\ \hline
D8 & Female & Korean-American & 25-29 & Entry level \\ \hline
D9 & Male & Asian & 25-29 & Intermediate level \\ \hline
\end{tabular}
\caption{Ideation session participants' basic demographic information. All questions (except job level) were free response.}
\Description{Table 1: Contains 9 observations (rows) and 5 features (columns). Features are ID, Gender, Race and Ethnicity, Age, and Job Level.}
\label{tab:codesignparticipants}
\end{table}
}

\newcommand{\intexamples}{
\begin{figure}
\centering
\begin{subfigure}[t]{.7\textwidth}
  \centering
  \includegraphics[width=.9\linewidth]{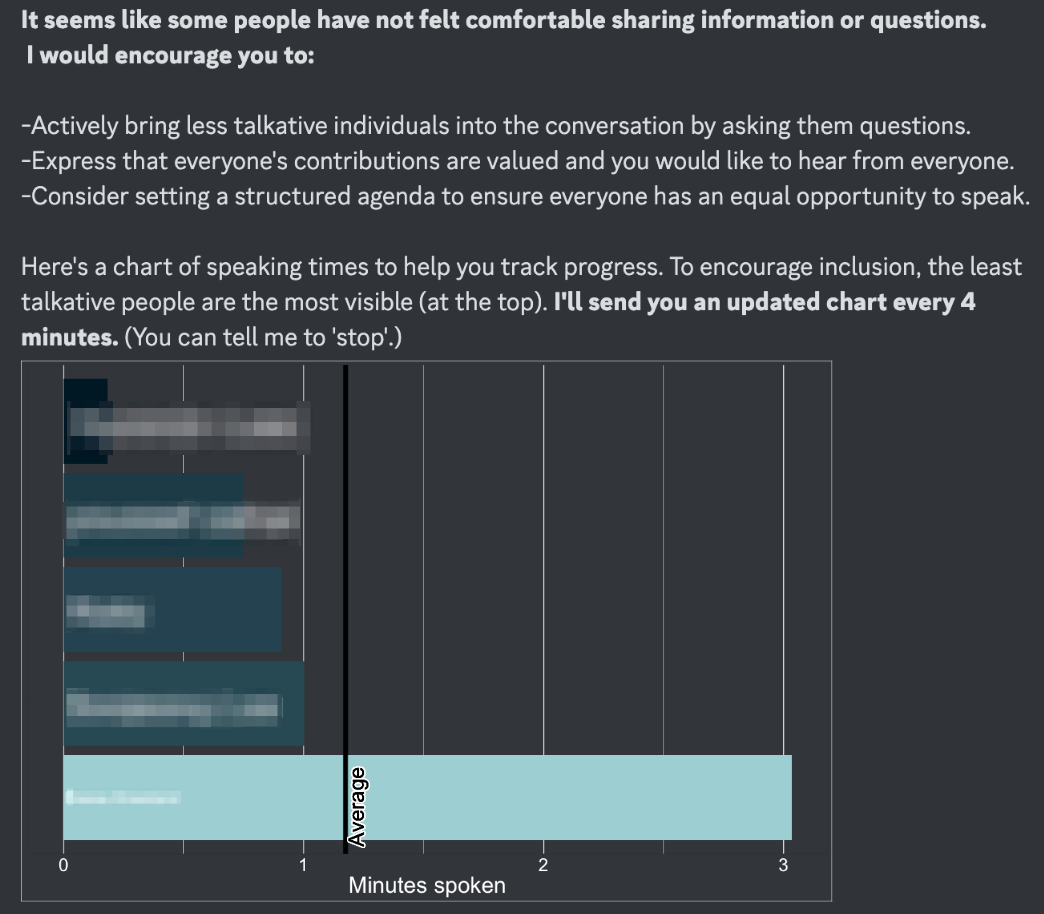}
  \subcaption{Host message example}
\end{subfigure}
\begin{subfigure}[t]{.8\textwidth}
  \centering
  \includegraphics[width=.9\linewidth]{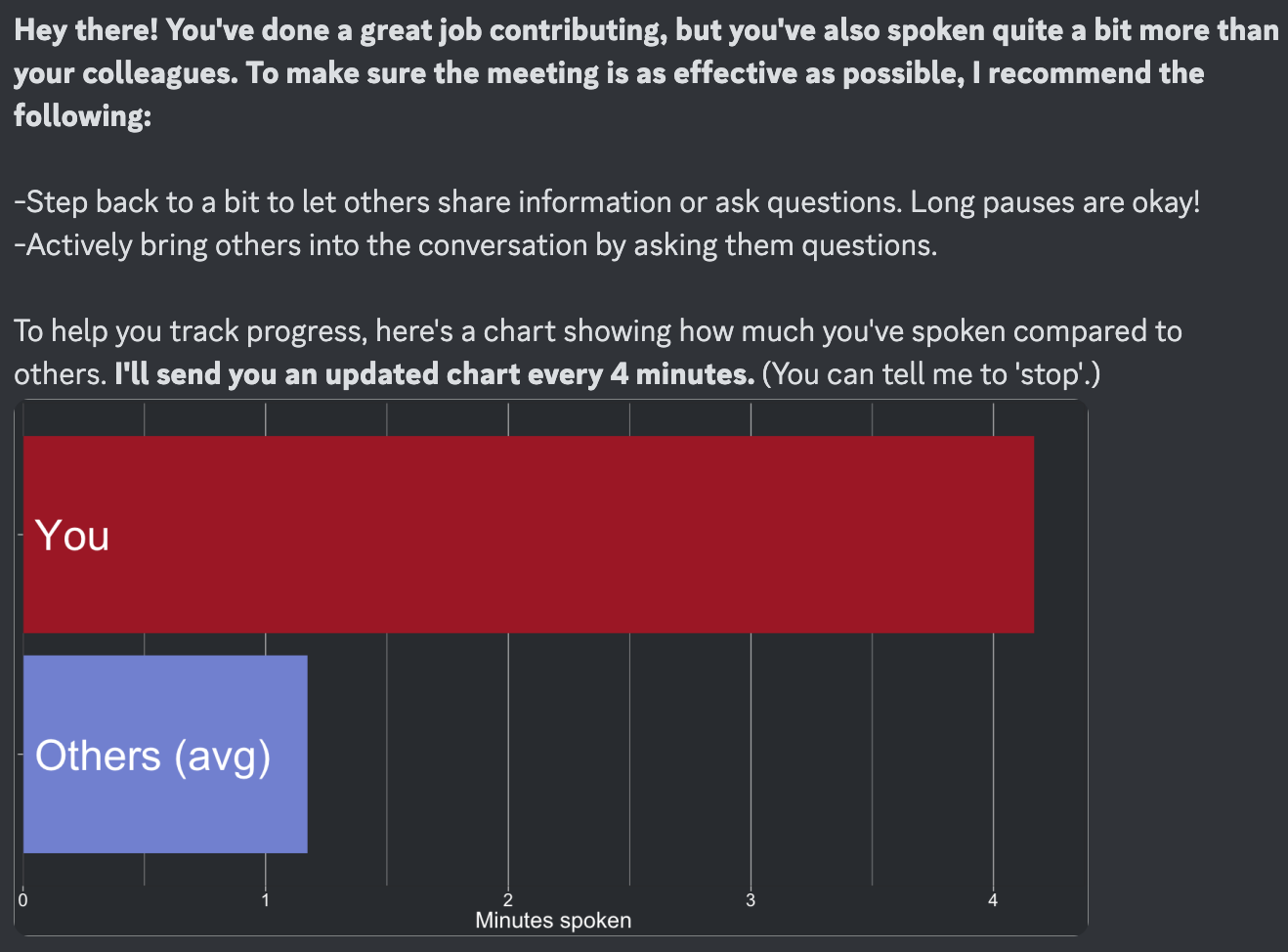}
  \subcaption{Over-participator message example}
\end{subfigure}
\caption{Examples of intervention messages from the virtual co-host. The virtual co-host notified participants of issues in the meeting, provided suggested actions, and included speaking time visualizations. Host-targeted visualizations made under-participators more visible by placing those who spoke least at the top and scaling name size inversely with speaking time.}
\Description{Figure 4: Two examples of the virtual co-host's intervention messages. The host message says some participants have not felt comfortable sharing info or questions, lists recommended actions like actively bringing less talkative individuals into the conversation, and concludes with a visualization showing each meeting participant's speaking time. The over-participator message tells the person they have spoken quite a bit more than their colleagues, and recommends they step back a bit to let others share information or ask questions. It concludes with a visualization showing the user's current time spoken and the average time spoken of other meeting participants.}
\label{fig:intexamples}
\end{figure}
}

\newcommand{\triggermessagecta}{
\begin{figure}[b]
\centering
  \includegraphics[width=.85\linewidth]{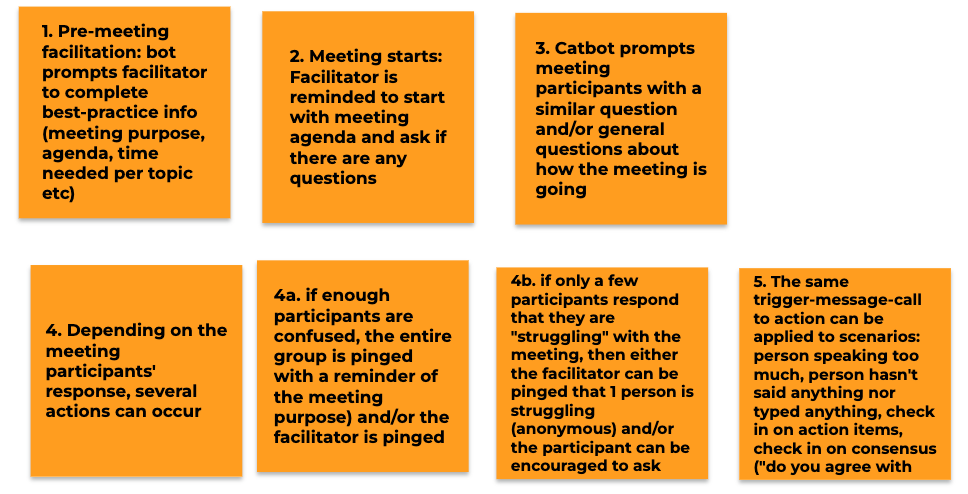}
  \caption{Partial screenshot of a participant's Jamboard, explaining their preferred virtual co-host features. Sticky notes 3-5 describe the ``trigger, message, call to action'' cycle, which we adapted to ``Observe, Ask, Intervene''.}
  \Description{Figure 2: 7 virtual sticky notes describing the ``trigger, message, call to action'' cycle. The ``trigger'' portion is not explicitly described in the sticky notes. The ``message'' portion is described as ``prompts meeting participants with... general questions about how the meeting is going.'' The ``call to action'' portion is described as ``if only a few participants respond that they are struggling with the meeting, then... the facilitator can be pinged that 1 person is struggling (anonymous).''}
\label{fig:triggermessagecta}
\end{figure}
}

\newcommand{\codesign}{
\begin{figure}[b]
\centering
  \includegraphics[width=.5\linewidth]{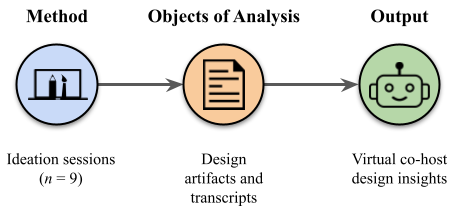}
  \caption{Summary of ideation session methods, analysis objects, and outputs.}
  \Description{Figure 1: Visual summary of the ideation session process. Method is 9 ideation sessions; objects of analysis are design artifacts and transcripts; output is virtual co-host design and implementation.}
\label{fig:codesign}
\end{figure}
}

\newcommand{\oai}{
\begin{figure}[t]
\centering
  \includegraphics[width=.8\linewidth]{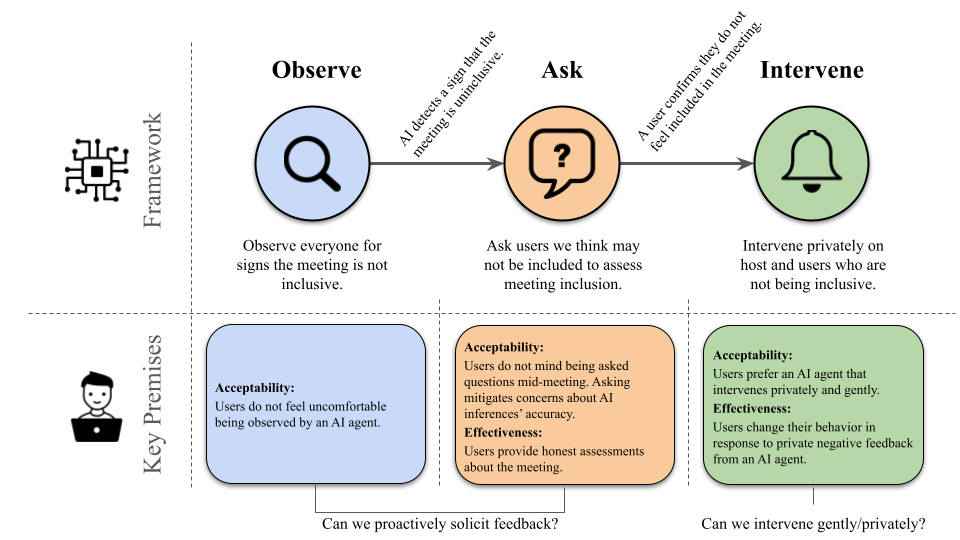}
  \caption{The ``Observe, Ask, Intervene'' (OAI) framework around which we designed our system, and the corresponding key premises about users we sought to test in each phase of the framework. These premises bear on the feasibility of our two major design insights: proactively soliciting feedback and intervening gently/privately.}
  \Description{Figure 3: A diagram overview of the OAI framework, key premises, and the design insights they pertain to. At the top is the OAI framework - Observe everyone for signs the meeting is not inclusive, Ask users to assess meeting inclusion, and Intervene privately on host and users who are not being inclusive. On the bottom are key premises. Observe: Acceptability: Users do not feel uncomfortable being observed by an AI agent. Ask: Acceptability: Users do not mind being asked questions mid-meeting. Asking mitigates concerns about AI inferences' accuracy. Effectiveness: Users provide honest assessments about the meeting. Intervene: Acceptability: Users prefer an AI agent that intervenes privately and gently. Effectiveness: Users change their behavior in response to the private negative feedback from an AI agent.}
\label{fig:oai}
\end{figure}
}

\newcommand{\labstudy}{
\begin{figure}[b]
\centering
  \includegraphics[width=.6\linewidth]{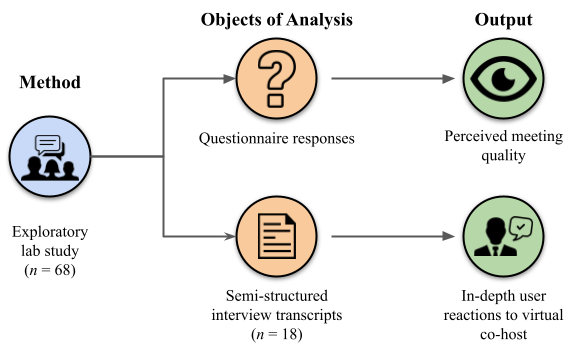}
  \caption{Summary of exploratory lab study methods, analysis objects, and outputs.}
  \Description{Figure 6: Visual summary of the exploratory lab study process. Method is exploratory lab study with 12 groups; objects of analysis are post-study questionnaire responses and 18 semi-structured interview transcripts; outputs are perceived meeting quality and in-depth user reactions to virtual co-host.}
\label{fig:labstudy}
\end{figure}
}

\newcommand{\cohostquestion}{
\begin{figure}[t]
\centering
  \includegraphics[width=.95\linewidth]{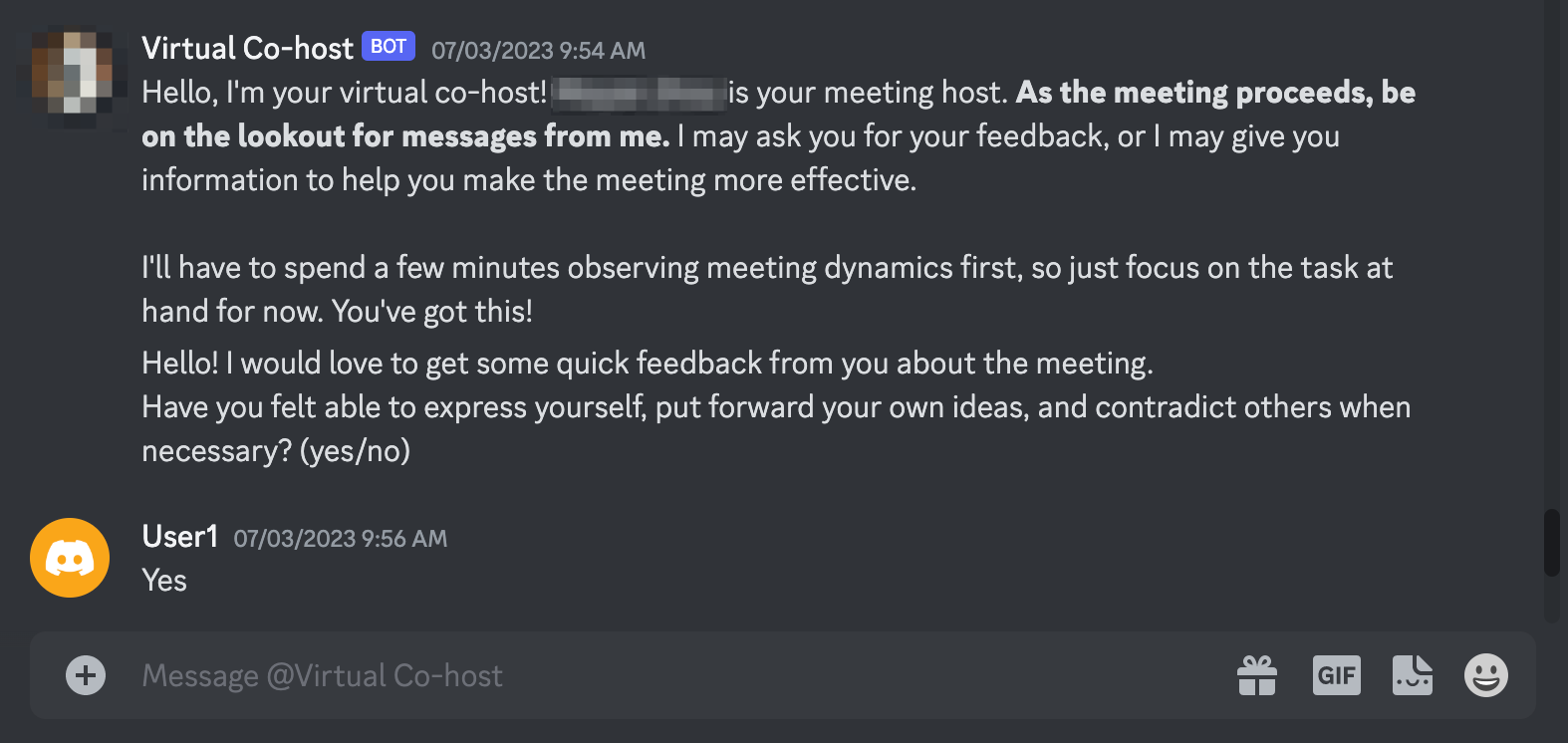}
  \caption{Screenshot of the co-host's intro message (sent to all participants shortly before start of the meeting) and first meeting assessment question (sent to under-participators several minutes later). Participants could respond to questions directly in text-based chat.}
  \Description{Figure 5: A series of messages in Discord's desktop interface. At the top, the virtual co-host introduces itself as the meeting host, tells the user to look out for messages from it, and tells them it has to observe meeting dynamics first. Message below show the virtual co-host asking for feedback from the user by asking them the first meeting assessment question. The user responds with ``Yes''.}
\label{fig:cohostquestion}
\end{figure}
}

\newcommand{\distraction}{
\begin{figure}[b]
\centering
  \includegraphics[width=.8\linewidth]{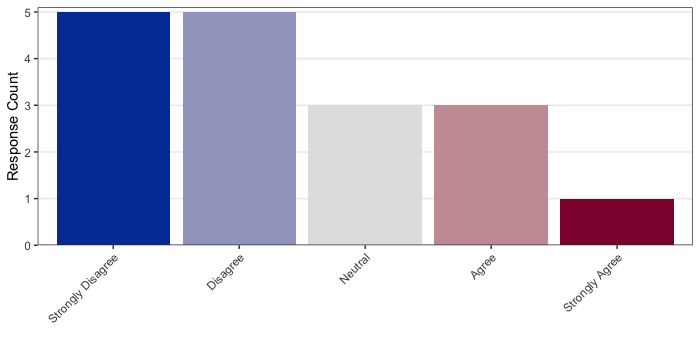}
  \caption{Agreement with the statement ``the virtual co-host was distracting'' by participants to whom the virtual co-host asked questions. }
  \Description{Figure 7: Histogram showing the occurrence count of each response. Strongly Disagree is 5, Disagree is 5, Neutral is 3, Agree is 3, and Strongly Agree is 1.}
\label{fig:distraction}
\end{figure}
}

\newcommand{\subjectiveratings}{
\begin{table}[]
\centering
\def\arraystretch{1.5}
\begin{tabular}{|l|r|r|r|r|r|r|r|}
\hline
\textbf{Construct} & \textbf{\textit{M} (control)} & \textbf{\textit{SD} (control)} & \textbf{\textit{M} (treatment)} & \textbf{\textit{SD} (treatment)} & \textit{\textbf{df}} & \textit{\textbf{t}} & \textit{\textbf{p}} \\ \hline
Communication & 8.45 & 3.37 &  \textbf{6.18} & 2.96 & 55.90 & 2.89 & **0.005 \\ \hline
Discussion Quality & \textbf{17.24} & 2.18 & 17.10 & 2.62 & 65.11 & 0.24 & 0.81 \\ \hline
Status Effects & 7.83 & 3.75 & \textbf{6.05} & 2.84 & 50.19 & 2.14 & *0.04 \\ \hline
Teamwork & 12.38 & 1.86 & \textbf{12.90} & 1.79 & 59.15 & -1.15 & 0.25 \\ \hline
\end{tabular}
\vspace{10pt}
\caption{Welch's two sample $t$-tests of user assessments on four of the five meeting quality constructs in Davison~\cite{davison_instrument_1999}. Participants who had the co-host present rated meetings better on all but one construct. 5-point Likert scale responses were summed within each construct to derive a corresponding value for each participant. Note that some of the constructs are worded positively while others are worded negatively~\cite[see~A.1]{davison_instrument_1999}, so the ``better'' mean value for each row is indicated in bold.}
\Description{Table 2: Contains t-test results for four meeting constructs. From left to right, the columns represent the construct, the mean response for control groups, the standard deviation for control groups, the mean response for treatment groups, the standard deviation for treatment groups, the degrees of freedom, the t-value, and the p-value of a two-sample Welsh's $t$-test.}
\label{tab:subjectiveratings}
\end{table}
}

\usepackage{xcolor}

\begin{document}

\title[Observe, Ask, Intervene]{Observe, Ask, Intervene: Designing AI Agents for More Inclusive Meetings}

\author{Mo Houtti}
\email{houtt001@umn.edu}
\affiliation{
  \institution{Department of Computer Science \& Engineering, University of Minnesota}
  \city{Minneapolis}
  \state{Minnesota}
  \country{USA}
}

\author{Moyan Zhou}
\email{zhou0972@umn.edu}
\affiliation{
  \institution{Department of Computer Science \& Engineering, University of Minnesota}
  \city{Minneapolis}
  \state{Minnesota}
  \country{USA}
}

\author{Loren Terveen}
\email{terveen@umn.edu}
\affiliation{
  \institution{Department of Computer Science \& Engineering, University of Minnesota}
  \city{Minneapolis}
  \state{Minnesota}
  \country{USA}
}

\author{Stevie Chancellor}
\email{steviec@umn.edu}
\affiliation{
  \institution{Department of Computer Science \& Engineering, University of Minnesota}
  \city{Minneapolis}
  \state{Minnesota}
  \country{USA}
}

\renewcommand{\shortauthors}{Mo Houtti, et al.}

\begin{abstract}
Video conferencing meetings are more effective when they are inclusive, but inclusion often hinges on meeting leaders' and/or co-facilitators' practices. AI systems can be designed to improve meeting inclusion at scale by moderating negative meeting behaviors and supporting meeting leaders. We explored this design space by conducting 9 user-centered ideation sessions, instantiating design insights in a prototype ``virtual co-host'' system, and testing the system in a formative exploratory lab study ($n=68$ across 12 groups, 18 interviews). We found that ideation session participants wanted AI agents to ask questions before intervening, which we formalized as the ``Observe, Ask, Intervene'' (OAI) framework. Participants who used our prototype preferred OAI over fully autonomous intervention, but rationalized away the virtual co-host's critical feedback. From these findings, we derive guidelines for designing AI agents to influence behavior and mediate group work. We also contribute methodological and design guidelines specific to mitigating inequitable meeting participation.
\end{abstract}



\received{12 September 2024}
\received[revised]{10 December 2024}
\received[accepted]{16 January 2025}

\maketitle

\section{Introduction}

Video conferencing (VC) platform adoption has exploded in recent years~\cite{iyengar_zooms_2020,sukhanova_zooming_2023}. Today, platforms like Zoom support large and ever-increasing amounts of group work across schools~\cite{stefanile2020transition}, workplaces~\cite{carradini2024evidence}, and government agencies~\cite{noauthor_about_nodate}. This shift to VC-mediated group work has renewed interest in HCI to innovate in and improve virtual meetings (e.g.,~\cite{samrose_meetingcoach_2021,houtti_all_2023,murali_affectivespotlight_2021, cutler_meeting_2021}).

An important yet challenging dimension for improving meetings is making them \textit{inclusive}---enabling all attendees to participate to the extent they desire. Inclusion in organizations is important for many reasons---it improves satisfaction~\cite{brimhall_mediating_2014}, performance~\cite{chen_does_2018}, and creativity~\cite{leroy_fostering_2022}. However, making meetings inclusive is hard work for meeting leaders and participants, so they often turn to co-facilitators for assistance~\cite{houtti_all_2023}. This is a good solution, consistent with prior work finding that facilitators enhance meeting outcomes~\cite{beranek_facilitation_1993, anson_experiment_1995, adkins_using_2003}. But facilitation also is hard work and many teams do not have the requisite training or funding to hire a facilitator~\cite{hayne_facilitators_1999}. 

We know from decades of research on group work that technology mediates our ability to do work well~\cite{olson2000distance}---and we argue that technology mediates some of the challenges of making VC meetings more inclusive. The benefits of inclusion translate to the VC meeting context~\cite{cutler_meeting_2021}, but making VC meetings inclusive is challenging because of sociotechnical barriers~\cite{houtti_all_2023}. Recently,~\citet{houtti_all_2023} found that the burden of facilitation falls on VC meeting leaders, who are often ill-equipped to run meetings equitably and do not have technical affordances to let them delegate their responsibilities. Taken together, we see a gap in the technology design of VC meetings, the pragmatic way they are managed, and the potential of technology to serve as a facilitator to make meetings more inclusive. This suggests a role for systems to act as a facilitator for inclusion to overcome issues of scalability and burden on attendees.

To explore this design space, we adopted the metaphor of a ``virtual co-host'' to facilitate more inclusive meetings, building on aforementioned work showing that better facilitation can improve meeting inclusion~\cite{houtti_all_2023} and other outcomes~\cite{beranek_facilitation_1993, anson_experiment_1995, adkins_using_2003}. We synthesize these insights in the VC space with ongoing work that uses AI as a tool for assisting in meetings~\cite{samrose_meetingcoach_2021} and as an agent for delegation. Using this metaphor as a starting point, we conducted ideation sessions to solicit feature ideas for an agent that could engage with meeting participants. Participants designed co-hosts to seek explicit feedback from meeting participants \textit{before} deciding to intervene on their behalf. We formalized this as an AI framework called ``Observe, Ask, Intervene'' (OAI). Participants also designed co-hosts to give negative feedback \textit{privately}, revealing an exciting opportunity for AI systems---to mitigate the social discomfort that occurs when humans communicate negative feedback to each other~\cite{abi-esber_just_2022,bond_reluctance_1987,stone_thanks_2015,jug_giving_2018,herrera_why_2018,eurich_right_2018}.


We tested these insights by prototyping a rule-based AI virtual co-host and conducting a formative evaluation through an exploratory lab study. The lab study included 68 participants across 12 groups of size 4-7. All groups completed a structured task through the same VC platform, with our virtual co-host present in 7 of the 12 groups. We collected in-depth qualitative feedback by interviewing 18 participants who interacted with the co-host, and quantitatively analyzed post-study questionnaires on meeting quality. We found that asking questions prior to interventions gave participants feelings of agency, was preferred over fully autonomous intervention, and was minimally distracting, but that hosts and over-participators rationalized away the virtual co-host's critical feedback. Despite the feedback's ineffectiveness, we also found that subjective assessments of meeting quality were better with the virtual co-host present.

Our findings inform the design of AI systems that seek to influence human behavior and/or mediate synchronous group work. We contribute:

\begin{itemize}
    \item details and operationalization of the ``Observe, Ask, Intervene'' (OAI) framework. By enlisting user judgment, OAI lets AI systems intervene without concerns about inaccuracy seen in prior work~\cite{samrose_meetingcoach_2021}. We show that OAI improves user agency, discuss how it follows from broader principles in human-AI interaction, and provide detailed design recommendations for OAI-based systems.
    \item evidence of a key limitation \textit{and} benefit of AIs as social actors~\cite{nass_computers_1994}: they exert little normative pressure. Based on this we discuss implications for designing \textit{effective} interventions, providing examples of how our system's chosen interventions could be altered to have the desired effects. We further highlight opportunities for leveraging AI to promote otherwise difficult feedback exchange in group contexts.
    \item methodological and design guidelines for future work on using AI systems to make meetings more inclusive. For example, we highlight key shortcomings of user-centered methods when designing such systems.
\end{itemize}

We present our work in two parts. We first focus on the ideation sessions and the prototype that emerged from them. We then cover our exploratory evaluation of the virtual co-host, including follow-up interviews with participants. We conclude by discussing implications for designing AI systems in and beyond the virtual meeting context.

\section{Related Work}

In this section, we outline the current state of sociotechnical solutions for VC meetings, other work informing the design of AI agents, and prior research on human-AI interaction.

\subsection{Sociotechnical Solutions for VC Meetings}

Video conferencing (VC) is understood to be limited in key ways compared to in-person interaction. For example, VC limits nonverbal communication~\cite{williams_working_2021, hills_video_2022, benford_shared_1996, houtti_all_2023}, which can reduce social cohesion~\cite{bonaccio_nonverbal_2016} and meeting satisfaction~\cite{hills_video_2022}, and increase psychological fatigue~\cite{bailenson_nonverbal_2021}.

Scholars and practitioners in and beyond CHI have explored social and sociotechnical solutions to many of VC's problems, often taking translucence-based~\cite{erickson2000social} approaches that make social information more visible to meeting attendees. In a series of experiments, ~\citet{hills_video_2022}, found that training students to employ ``video meeting signals''---a simple set of gestures used to communicate common feelings---increased meeting satisfaction. In an empirical study, ~\citet{nguyen_multiview_2007} found that non-traditional VC setups that preserve the ability for spatial referencing improved the formation of intragroup trust.~\citet{langner_eyemeet_2022} found that employing a joint attention eye-tracking system improved meeting participants' focus. Other solutions have leveraged AI. ~\citet{murali_affectivespotlight_2021} implemented AffectiveSpotlight, which overcomes the difficulty of gauging audience mood by identifying and spotlighting the most expressive person in the meeting at intervals. Samrose et al.'s MeetingCoach~\cite{samrose_meetingcoach_2021} uses AI to detect various behavioral cues in meetings, and consolidates them into a post-meeting dashboard that was shown to improve understanding of meeting dynamics. Many commercial systems (e.g.,~\cite{noauthor_airgram_2023, noauthor_firefliesai_2023,noauthor_rewatch_2023}) use AI to perform difficult and helpful administrative tasks such as transcription, note-taking, topic tracking, and/or summarization. 

Prior research suggests that, while useful, translucence-based approaches are inherently limited. For example, ``nudging''---influencing humans' behavior by altering how information is presented---has been shown to encourage prosocial behavior in a range of contexts~\cite{marteau_judging_2011,szaszi_systematic_2018,schneider_digital_2018}, but~\citet{hummel_how_2019} find in a quantitative review that nudges have small effect sizes. Informational approaches rely on meeting attendees to decide appropriate actions based on the information presented, but they do not attend to strong non-informational (often social) incentives that can influence behavior in counter-productive directions---e.g., structural inertia~\cite{hannan_structural_1984}, peer pressure~\cite{connor_peer_1994}, or cognitive dissonance~\cite{harmon-jones_introduction_2019}. AI agent-based approaches could overcome these limitations by taking a more active role in meetings, but existing work on designing AI agents for behavior change has mostly been limited to single-user contexts---e.g., encouraging regular breaks~\cite{reeder_breakbot_2010} and self-reflection~\cite{kocielnik_designing_2018}. We extend this work by exploring unique challenges (e.g., managing peer social pressure) that arise when designing AI agents to mediate synchronous group work.

We specifically explore how AI agent-based systems could address a common VC problem surfaced by~\citet{houtti_all_2023}, who found in an interview study that VC meetings rely too much on meeting leaders to make meetings more inclusive. They suggest two mechanisms for mitigating this issue in VC meetings: \textit{``transferring control from meeting leaders to technical systems or other attendees and helping meeting leaders better exercise the control they do wield.''} We used the virtual co-host metaphor to explore the design space of the former recommendation, by mirroring the social role of a person to whom meeting leaders often transfer control~\cite{houtti_all_2023, anson_experiment_1995, adkins_using_2003}. Our virtual co-host assumes joint responsibility over meeting inclusion by reminding hosts to facilitate more inclusively when needed, giving them information to help them do so, and giving other meeting participants direction on how to be more inclusive.

\subsection{Designing Automated Agents}

Extensive research has explored the design and use of automated (AI and non-AI) agents for various purposes.~\citet{liao_what_2016} prototyped a personal assistant to help employees find work-related information and evaluated it in a field study. They found, among other things, that interrupting users with proactive interaction is risky, especially for those with busy work schedules.~\citet{kocielnik_designing_2018} explored the use of a conversational agent for workplace self-reflection, highlighting user preferences and trade-offs between voice- and chat-based interaction.~\citet{fitzpatrick_delivering_2017} employed a conversational agent to administer cognitive behavioral therapy and found that it reduced feelings of depression.~\citet{schroeder_pocket_2018} did the same for dialectical behavioral therapy with positive results.~\citet{dmello_toward_2007} demonstrated that a conversational AI could tutor students.

Other work has shown how automated agents can influence behavior, delineating both their capabilities and limitations.~\citet{vollmer_children_2018}, for example, found that humanoid robots' answers to a question did not affect how adult humans answered, though \textit{other humans'} answers did. This suggests that, unlike humans, automated agents cannot exercise subtle peer pressure. Conversely,~\citet{bickmore_randomized_2013} found that a conversational agent was more effective than a pedometer in encouraging people to walk more, demonstrating the ability of automated agents to change behavior more effectively than passive informational approaches. Like~\citeauthor{bickmore_randomized_2013}, we explore the use of a proactive automated agent to change behavior, this time in the group work context of meetings. Virtual meeting inclusion often falls on meeting hosts~\cite{houtti_all_2023}, so we use the ``co-host'' metaphor to solicit designs for an automated agent that assumes some of this responsibility.

\subsection{Human-AI Interaction}

AI systems are commonly implemented to be responsive to user interaction. For example, the popular ``human-in-the-loop'' paradigm involves users to iteratively improve algorithmic decision-making. Interaction can take many forms: controlling the learning process~\cite{mosqueira2023human}, providing training data~\cite{wu2022survey}, annotating and labeling raw data~\cite{monarch2021human}, fact-checking against misinformation~\cite{demartini2020human}, and much more~\cite{zanzotto2019human, fugener2021will}.

As the number of user-facing AI applications has increased, a growing body of work has looked to inform how we should design human-AI interaction at the interface level.~\citet{amershi_guidelines_2019}, for example, propose 18 general design guidelines for human-facing AI technologies. Among them is \textit{``Enabl[ing] the user to provide feedback indicating their preferences during regular interaction with the AI system.''} Schmidt~\cite{schmidt_interactive_2020} similarly focuses on the importance of control in human-AI interaction, arguing that \textit{``being able to determine what should happen is strongly related to self-determination and freedom of choice, and is ultimately a basis for feeling safe.''}~\citet{schmidt_intervention_2017} propose ``intervention user interfaces'' allowing humans to intervene in AI processes with immediate effect.~\citet{lai_human-ai_2022} propose a paradigm where users circumscribe ``trustworthy regions'' of action that can be conditionally delegated to AI systems.

Our work builds on this. Prior work and our ideation sessions led us to test a system that defers to user judgment when deciding how and whether to intervene. Therefore, our exploratory evaluation of the system let us test claims made by~\citeauthor{schmidt_intervention_2017} and others about the importance of agency in designing interactive AI systems.

\section{Part 1: User-Centered Ideation}

Recall that our work is split into two parts: user-centered ideation sessions and an exploratory lab study. We begin with the former, outlining the methods, findings, and the system they led us to prototype.

\subsection{Ideation Session Methods}

\codesign

We conducted 9 ideation sessions over Zoom (Figure ~\ref{fig:codesign}) with a semi-structured design activity. (Basic demographic information is reported in Table~\ref{tab:codesignparticipants}). Participants had all engaged in group work via VC platforms through their employment or other means. We received approval for our study from our institution's IRB. All participants were invited to participate by a research team member through email, either because they were direct personal connections or because they had participated in a previous study in our lab and had consented to be contacted for follow-up research studies. Participants were compensated with \$20 Amazon gift cards. 

\codesignparticipants

The ideation sessions let us explore design possibilities and needs in detail. Recall that in the prior work (e.g.,~\cite{houtti_all_2023, anson_experiment_1995, adkins_using_2003}, see Related Work for more), better facilitation improves meeting outcomes. Therefore, we began each session by presenting the virtual co-host metaphor,  explaining that its goal was to reduce inequitable barriers to participation. To do this, it could behave like any other (human) participant in the meeting---i.e., it could interact with other attendees through text, audio, or any of the platform's other affordances like screen-sharing, but could not modify the platform's UI. We set these constraints to ensure participants focused their ideation on the agent rather than improvements to video conferencing platform UIs, in accordance with our research goals as informed by prior work. The rest of the session was then split into two parts:

\begin{itemize}
    \item \textbf{Design activity:} We asked each participant to design a virtual co-host that could intervene to mitigate inequitable barriers to participation in their workplace meetings. We asked each participant contextual questions to keep them moving towards a final design (e.g., \textit{``which of the two approaches you proposed do you think is more compelling?''}) or to clarify the reasoning behind design decisions (e.g., \textit{``how do you see this feature reducing inequity?''} or \textit{``why should the co-host send this information privately instead of publicly?''}). ($\sim$30 min)
    \item \textbf{Presentation:} We gave participants time to sketch their final idea on a Google Jamboard. They then presented their idea to us, and we asked follow-up questions as needed to understand the motivations behind specific design decisions. ($\sim$15 min)
\end{itemize}

Two of the authors conducted the ideation sessions. Sessions were recorded and transcribed for future reference. We adopted a collaborative design thinking~\cite{micheli_doing_2019} approach to analyze artifacts and develop a final design. From the Jamboard artifacts, the two authors identified key design insights, described in the following subsection. Building directly on features designed by participants, they collaboratively brainstormed how to design feasible iterations that would instantiate the key design insights, could plausibly alter meeting behavior, would be measurable in simulated lab study meetings, and could give us useful information about designing AI agents for improving meeting inclusion. These considerations were balanced by synthesizing participant ideas, design insights, and insights from prior research.

\subsection{Ideation Session Findings}

From our ideation sessions, we gleaned several insights about \textit{when} and \textit{how} a virtual co-host should intervene in meetings. We outline these design insights next and, where relevant, we highlight how prior work shaped our thinking on them.

\subsubsection{When? After asking for explicit user input}

There was a ubiquitous preference for interactive over fully autonomous AI. Users' co-host ideas demonstrated a desire to retain some human control over the co-host's behavior. For example, D6 designed a co-host to ask people \textit{``what is the purpose of this meeting?''} before deciding whether to intervene on their behalf. D7 designed a co-host to provide nudges based on behaviors the user says they want help with. Other examples include transcribing meeting content when asked (D1), pinging the host when asked (D9), delivering questions anonymously when asked (D4), and nudging the current speaker when someone indicates they would like to speak (D3, D5).

Interactivity could plausibly reduce mistrust in the system's inferences. Participants in~\citet{samrose_meetingcoach_2021} were skeptical about AI's ability to accurately assess their internal states based on external cues: (\textit{``I don't feel that AI is good enough at sentiment.''} \textit{``I have some doubts on whether this can be done accurately, given how subjective some of this information would be.''}~\cite{samrose_meetingcoach_2021}) Deferring to humans has been used as a solution in other, high-stakes domains where users are worried about AI system accuracy~\cite{kawakami_improving_2022, smith_keeping_2020, binns_human_2022}. Making our co-host rely on user input would let us test whether similar solutions can improve trust in AI systems for VC meetings.

\triggermessagecta

The co-host should be \textit{proactive} in soliciting user input. D6 designed a co-host that followed what they called the ``trigger, message, call to action'' cycle, as shown in Figure~\ref{fig:triggermessagecta}---a process where the virtual co-host is triggered into action by automated inferences, messages affected attendees to verify its inferences, then provides a targeted call to action based on feedback. Prior work suggests many VC features are underused because meeting participants must know of them and seek them out~\cite{houtti_all_2023}. We reasoned that the onus should not be on users to always prompt the virtual co-host's actions or it would be an underused feature. We therefore adopted D6's approach, and renamed it ``Observe, Ask, Intervene'' (OAI).

\subsubsection{How? Non-intrusively and gently}

Ideation participants imagined lightweight, non-intrusive, and friendly interventions. D6, for example, proposed a feature that communicates whether anyone is struggling with the meeting content (Figure~\ref{fig:triggermessagecta}). D7 proposed a virtual co-host that alerts the host when they move to a new topic without letting others chime in. D8 noted that feedback messages should be \textit{``something friendly''}. D2 and D6 suggested speaking time as a useful baseline metric for understanding whether everyone can participate, and noted that visualizations could make feedback easier to digest.

\subsubsection{How? Privately}

Implicit in participants' preferences for ``friendly'' interventions was a desire to avoid creating social discomfort. This manifested more explicitly when they articulated why the co-host should intervene \textit{privately}. When asked to whom speaking time visualizations should be displayed, D6 said: 

\begin{quote}
\textit{``It's better just to have it for the facilitator, or maybe just for the person who's speaking too much... I wouldn't share it with the whole group, I would just share it with the people whose behavior you're trying to change, which in this case sounds like those who are speaking too much. ''} 
\end{quote}

She predicted that public interventions would make over-participators \textit{``anxious, and counting the seconds as they're speaking''} and wanted to avoid creating discomfort. D3, who designed a co-host to deliver text-based nudges to the current speaker, also indicated that feedback should be targeted at the person whose behavior we are trying to change. D2 expressed that public interventions might even make under-participators uncomfortable by highlighting that \textit{``this person hasn't spoken for a little while.''} 

This was valuable given that receiving negative feedback from \textit{human} peers is a frequent source of social discomfort~\cite{abi-esber_just_2022,bond_reluctance_1987,stone_thanks_2015,jug_giving_2018,herrera_why_2018,eurich_right_2018} even when the feedback is given privately. \textbf{Combining prior work with our participants' requests revealed an exciting opportunity: to deliver critical feedback \textit{without} social discomfort by having an AI agent deliver feedback instead of a human.} Prior work has shown that publicly shared speaking time visualizations during meetings lead over-participators to speak less~\cite{dimicco_impact_2007}. Supplementing visualizations with private feedback and suggestions from an automated agent---as opposed to displaying impersonal labeled visualizations to the group---could make their already demonstrated effects stronger and more effective, while letting us retain the social comfort of private interventions that participants valued.

\section{Part 2: Virtual Co-host Implementation and Exploratory Lab Study}

We extracted two major design insights from the ideation sessions. A virtual co-host should (1) \textit{proactively} solicit feedback from select meeting participants to inform its interventions and (2) intervene with non-intrusive, gentle, private feedback. Participants suggested concrete ways in which these insights could be instantiated in a system. Guided by their suggestions, we implemented a prototype rule-based virtual co-host system and evaluated it in an exploratory lab study. To satisfy the first design insight, we organized the virtual co-host's design around the ``Observe, Ask, Intervene'' (OAI) framework adapted from D6's ``trigger, message, call-to-action'' idea. To satisfy the second design insight, we selected interventions for the ``Intervene'' phase that were non-intrusive, gentle, and private.

Our evaluation focused on whether key premises justifying our design insights would be borne out empirically at each stage of the virtual co-host's behavior in OAI. Specifically, we sought to determine whether each phase of OAI would be \textit{acceptable} to users and \textit{effective} at helping the system achieve its functional purpose. (We only considered acceptability for ``Observe'' given that AI inference's effectiveness is already well studied~\cite{murali_affectivespotlight_2021,samrose_meetingcoach_2021}.)

The first design insight's feasibility relied on user reactions in the ``Observe'' and ``Ask'' phases. To proactively solicit feedback, we must be able to \textbf{observe} for signals of a non-inclusive meeting, and react to those signals by \textbf{asking} questions mid-meeting. Users should provide honest information in response. The second design insight's feasibility relied on user reactions in the ``Intervene'' phase. To be worthwhile, private and gentle interventions should lead to changes in behavior that improve the meeting. Each stage in this process must independently be \textit{acceptable} to users as well, if not preferable. In Figure~\ref{fig:oai}, we provide a complete summary of OAI, key premises about each phase we sought to elucidate, and their relation to our two major design insights.

Our design insights encompass an expansive design space---e.g., we could think of many plausible interventions that would be non-intrusive, gentle, and private. We therefore opted for a primarily qualitative exploration, supplemented by descriptive quantitative data. This let us go beyond simply understanding the effects of our \textit{particular} implementation. Knowing both if and \textit{why} key premises underlying OAI are (not) correct would let us develop a rich understanding of the \textit{overall} design space, which is preferable for this kind of formative study.

\oai

Our implementation instantiated design insights only as needed to adequately evaluate them. Similar to a ``Wizard of Oz'' experiment~\cite{dahlback1993wizard}, we selectively withheld information about how the system really works to understand how users might react to a robust version of the system. We note these decisions throughout the system description, which is organized around OAI's three phases.

\subsection{Virtual Co-host Implementation}

\subsubsection{Observe}

\hfill\\The virtual co-host begins each meeting by introducing itself and letting everyone know who the host is. It tells users it will be ``observing conversational dynamics'' and may reach out later in the meeting (Figure~\ref{fig:cohostquestion}). To inform its Ask and Intervene phases, the virtual co-host observes and records each participant's cumulative speaking time at 1-second intervals. While the virtual co-host only observes speaking times, we chose to not have it tell meeting participants what information it collected. We left the phrase ``observe conversational dynamics'' vague because we wanted to understand whether participants would feel a ``watching-eye'' effect~\cite{hu_dark_2023} when being observed by the co-host if they did not know what info it was collecting (e.g., tone, sentiment, interruption patterns, etc.), and whether adding the Ask phase would mitigate or exacerbate those concerns.

For the purposes of our study, we used over-inclusive rule-based triggers for the Ask phase. We were most interested in the efficacy of the Ask and Intervene phases and did not want the Observe phase to act as a bottleneck. The goal was to ensure the virtual co-host would wait for some minimal level of inequality to manifest but \textit{always} proceed to the Ask phase even if no substantial inequality was detected. Future iterations could use more advanced AI to select effective Observe rules; however, we did not see this as necessary for testing the concept in a controlled lab environment. Therefore, the virtual co-host moves to the Ask phase if:

\begin{enumerate}
\item at least 8 minutes have elapsed, and a non-host participant has spoken more than twice the average or less than half the average, \textbf{or}
\item half the meeting time has elapsed.
\end{enumerate}

The latter rule ensured that the Ask phase would always be triggered by minute 15 of each 30-minute lab study meeting. This would ensure that under-participators had a chance to answer the virtual co-host's questions and that enough time would be left for interventions to affect the meeting.

\cohostquestion

\subsubsection{Ask}

\hfill\\The virtual co-host identified anyone with below-average speaking time as an under-participator. We, again, used over-inclusive criteria so more participants would receive the virtual co-host's questions (Figure~\ref{fig:cohostquestion}), letting us understand user reactions to being asked for feedback by an AI co-host. Under-participators were then asked:

\begin{enumerate}
\item Have you felt able to express yourself, put forward your own ideas, and contradict others when necessary?
\item Have you felt inhibited from participating in the discussion because of the behavior of other meeting members (other than the host)?
\item Is there any feedback or advice you'd like me to anonymously pass on to the host?
\end{enumerate}

Questions 1 and 2 were adapted from Davison's revalidated meeting assessment instrument~\cite{davison_instrument_1999} to ensure they captured common meeting problems, selecting the questions that closely matched our outcomes. Participants could respond to the virtual co-host by typing their answers in a direct chat message. The first two questions required a yes or no answer. For the third, participants could either type a free-response message or reply with ``no''. When given unsupported input, the virtual co-host would reply that it did not understand, prompting participants to try again.

A ``no'' answer to Question 1 activated the host intervention. A ``yes'' answer to Question 2 activated both the host and over-participator interventions. Note that we decided to include the host intervention if \textit{any} behavioral change was needed since the host is central to how a meeting runs~\cite{sauer_ties_2015,schuleigh_effects_2021, schuleigh_enhancing_2019, lehmann_critical_2017, jakobsson_energizing_2021}. In some meetings, multiple people received the questions from the virtual co-host, in which case any single negative response activated the corresponding intervention(s). 

To minimize distraction, we made the virtual co-host easy to ignore in the Ask stage; if the user does not reply, it does nothing. The co-host also provides minimal detail about its interventions; after receiving a negative assessment, it lets the user know it will intervene but does not tell them how. We reasoned that an OAI system for meetings would need to be minimally distracting, and should therefore operate with minimal oversight (like a human co-host would). Aligning our system's design with this principle let us understand whether a realistic level of user interaction---quickly soliciting users' perspectives, rather than giving them full insight into and control over the virtual co-host's behavior---would give us the expected benefits of interactivity and be acceptable to users.

\subsubsection{Intervene}

\hfill\\If the host intervention was activated, the virtual co-host sent the host a message indicating a meeting problem and recommended strategies for facilitating more inclusively. The message varied based on which question prompted the intervention, but always contained a bar chart with information about each member's time spoken, in accordance with D2 and D6's ideas.~\citet{houtti_all_2023} found that under-participators often become ``invisible'' in virtual meetings, thereby favoring over-participators being engaged in conversation. We therefore designed this visualization to make under-participators the most visible to the meeting host (Figure~\ref{fig:intexamples}a). Hosts occupy a special role that often requires them to speak more than the average person in the meeting. Therefore, the visualization did not include their speaking time, as we did not want to encourage the facilitator to step back from their responsibilities.

\intexamples

The virtual co-host considered the person who spoke the most an over-participator. This meant an over-participator might be identified even in a highly equal meeting; we once again selected over-inclusive criteria that were appropriate for our study objectives (understanding reactions to the virtual co-host's interventions). We also reasoned that an over-inclusive system would trigger on potential false positives (over-participation identified in a good meeting), allowing us to explore the acceptability of soliciting feedback when a meeting is going well. To catch situations with multiple over-participators, any other person who spoke more than twice the average was also considered an over-participator. This rule was never triggered, so each meeting had a single over-participator designated.

If the over-participator intervention was activated, the virtual co-host told over-participators they should let others contribute more and gave suggestions for improving their behavior. Unlike hosts, over-participators received different information and visualizations, about their speaking time relative to the \textbf{average} of all other non-host meeting participants (Figure~\ref{fig:intexamples}b). We reasoned that users frustrated with an over-participator were unlikely to feel comfortable with them having privileged access to every participant's exact speaking time. Doing so might also give them information with which to criticize under-participators for not contributing---a dynamic we would not want any system to encourage. 

Targets of the virtual co-host's interventions received an updated visualization every 4 minutes to encourage continued awareness of their (non-)compliance with the norms it expressed. We extensively piloted the visualizations to make sure they were visually salient and easy to understand. Unlike under-participators in the Ask stage---who would stop receiving messages if they ignored the virtual co-host---those receiving interventions had to tell the co-host to ``stop'' if they wanted to stop receiving messages from it. To prevent interruptions and ensure receivers did not miss the virtual co-host's messages, messages were queued and sent after the receiver's microphone had been inactive for at least 5 seconds.

\subsection{Technical Details}

The virtual co-host was implemented as a Discord bot using Discord.py. Discord is a free video chat system with over a hundred million users~\cite{noauthor_about_nodate}. It is often used for student project teams, course spaces, and gaming. The virtual co-host used Selenium and custom Javascript to track participants' speaking times in real-time by monitoring elements in Discord's web browser interface. It communicated with participants through direct messages.\footnote{Discord displays direct message notifications through a pop-up bubble on a sidebar, showing who sent the message and how many messages were sent. The user can click the bubble to open a direct message interface with the sender. This is similar to clicking to open chat in other VC software. When finished, they can go back to the meeting interface by clicking the meeting server's bubble in the same sidebar.} In addition to text, messages contained visualizations generated with R/ggplot2. Images are sent in their full dimensions through Discord (not as thumbnails or downloads like in other VC software). Participants in a pilot study meeting reported no difficulty noticing and seeing the co-host's visualizations.

\labstudy

\subsection{Exploratory Lab Study Methods}

Students at several US-based institutions were invited to sign up to participate in our study, alone or in groups of up to 3. Our institution's IRB reviewed and approved this study. Recruitment was done through in-person class announcements, a departmental research pool, and emails to institutional/departmental mailing lists. Experiment sessions had 4 to 7 participants. Each participant was compensated \$20 for completing the study. If participants signed up in a group, they were guaranteed to be in the same session. This setup ensured that in every session, most participants knew at least some participants, but no participant knew everyone---better approximating many group work dynamics.

Our study used a between-subjects design~\cite{CHARNESS20121}, which satisfied our research objectives while minimizing demand on participants' time. This approach let us reliably recruit enough participants to obtain deep qualitative insights (our primary objective) and quantitative data to supplement those insights.

We assigned groups to conditions randomly, except in cases where we needed to maintain balance between conditions. For example, we manually assigned two groups to the treatment condition following three random assignments to the control condition in a row. We started with 5 control group sessions ($n=28$) and 5 treatment group sessions ($n=29$). The virtual co-host was only present in treatment groups; control groups had no exposure to the system whatsoever. In the 5 treatment sessions, the virtual co-host's visualization- and feedback-based interventions were only triggered twice. (One intervention was correctly triggered, while the other resulted from a participant input error.) In another meeting, the host received positive free-form feedback passed anonymously from another participant, but no visualizations or guidance from the virtual co-host. It became apparent that gathering sufficient data on the system's interventions would not be feasible without a significantly larger sample size. Therefore, in two additional treatment groups, researchers manually triggered the interventions (minus the free-form feedback) regardless of under-participators' meeting assessments ($n=11$, bringing the total treatment group to $n=40$). This gave us more data with which to develop a rich understanding of user reactions to the ``Intervene'' phase.

Each session was structured as follows:

\begin{itemize}
    \item \textbf{Icebreaker}: Two researchers facilitated a short icebreaker activity. Participants were asked to share their names and one thing they enjoy doing.
    \item \textbf{Solo Activity}: Participants were given a worksheet to complete the Lost at Sea activity~\cite{biech_pfeiffer_2007}, a group work task used in prior work (e.g.~\cite{dugan_task_2002, chen_groupthink_1996,littlepage_effects_1997, hu_fluidmeet_2022}). In this activity, participants are given a hypothetical scenario where they are on a lifeboat in the sea and asked to rank 15 items in order of importance for survival. We selected this task in part because it does not \textit{require} equal participation (and could be completed alone). Participants completed this activity solo for 10 minutes to allow everyone adequate time to come up with their own ideas. (Timed, 10 minutes.)
    \item \textbf{Group Activity}: A meeting host was selected at random and was instructed to lead the process of producing a group answer for the Lost At Sea activity. \textbf{In treatment groups}: the virtual co-host joined the meeting and introduced itself (Figure~\ref{fig:cohostquestion}). The facilitating researcher alerted participants to the co-host's presence to ensure they were aware of it and how it communicated. We wanted to see how participants would react to the virtual co-host, so we did not give any instructions on how they should respond to its messages. (Timed, 30 minutes.)
    \item \textbf{Post-study Questionnaire}: Participants were given a Qualtrics survey containing Davison's meeting assessment instrument~\cite[Section~A.1]{davison_instrument_1999}, which assesses meetings on five constructs: Communication, Discussion Quality, Status Effects, Teamwork, and Efficiency\footnote{Our survey erroneously omitted the last two items in the instrument (E1 and E4), both of which pertain to the Efficiency construct. We, therefore, omit the Efficiency construct from our statistical results. Excluding this construct does not change the overall interpretation of our findings. Our analysis of questions E2 and E3 further supports this interpretation and is included in a Supplemental table~\ref{tab:suppmain}.}. People in treatment groups assessed how distracting the virtual co-host was on a 5-point Likert scale.
    \item \textbf{Semi-structured Interviews}: A subset of participants who experienced the co-host's Ask or Intervene phases were invited for follow-up interviews (10 under-participators, 3 over-participators, and 5 hosts; $n=18$), where we solicited in-depth reactions to the virtual co-host. Interviews lasted on average about 13 minutes.
\end{itemize}

\textbf{Analysis:} We analyzed follow-up interviews using inductive thematic analysis. We generated open codes for each unique participant statement, conducted axial coding to group-related codes, and concluded with selective coding to identify major themes and outcomes. We recorded audio and video of the group activities to assist in analysis. Participants were notified in the consent form and before the icebreaker that parts of the study would be recorded. To supplement the qualitative analysis, we conducted statistical tests on questionnaire responses. We report descriptive statistics where appropriate.

\subsection{Exploratory Lab Study Results}

We organize our findings by OAI's three steps. Recall that we wanted to understand whether and how a virtual co-host could be designed to be both \textbf{acceptable} (i.e., users like the system) and \textbf{effective} (i.e., it works as intended). Therefore, we articulate how our findings relate to acceptability and effectiveness. We address only acceptability for ``Observe'' since the effectiveness of AI inference in VC meetings is already demonstrated by prior work (e.g.,~\cite{murali_affectivespotlight_2021,samrose_meetingcoach_2021}). 

Each interview participant's anonymous ID consists of a letter---H, U, or O indicating they were a host, under-participator, or over-participator. For example, H1 was a host and U1 was an under-participator, though they may not have been in the same meeting. When relevant, we specify in text that participants were in the same meeting. We lightly edited participant quotes for brevity and clarity.

\subsubsection{\textbf{\underline{\large Observe \normalsize}}}\hfill\\

\noindent \textbf{\textit{\underline{Acceptability:}}}

\hfill\\\noindent\textit{\textbf{No negative effects of being observed by the co-host.}}

In interviews, participants reported no negative feelings about the virtual co-host's presence. The virtual co-host joined the meeting as a participant---it had its own user tile just like other meeting participants---and it sent an introductory message telling participants it would ``observe conversational dynamics.''  U10, for example, recalled feeling that \textit{``it was just kind of there''} and that they did not think about it until it messaged them. U6 said \textit{``my focus wasn't entirely on it.''} Our results connect to prior work that the experience of a ``watching-eye'' effect is context-dependent~\cite{hu_dark_2023}, and show that AI agents may not have this negative effect in virtual meetings even if they are visually presented as attendees with their own user tiles.

\subsubsection{\textbf{\underline{\large Ask \normalsize}}}\hfill\\

\noindent\textbf{\textit{\underline{Acceptability:}}}

\hfill\\\noindent\textit{\textbf{Engaging with the virtual co-host was generally positive.}}

Participants indicated that receiving questions from the virtual co-host was a generally positive experience. U5 appreciated that the co-host's questions helped validate their negative feelings about the meeting. \textit{``I was already feeling frustrated, so it was good to \textbf{validate} that.''} Knowing that the co-host would share the feedback did not change things. U6 guessed that giving the virtual co-host a negative response would cause it to message the host, but recalled that this made them \textit{``not particularly uneasy; maybe 10\%.``}

\hfill\\\noindent\textit{\textbf{Engaging with the virtual co-host was minimally distracting.}}

Participants' subjective evaluations of distraction also showed a willingness to engage, and some participants were not bothered by the co-host.  However, a few participants found the virtual co-host distracting and worried that if they responded negatively, \textit{``it's going to talk to me more.''} (U3) In the post-study questionnaire, participants in meetings with the virtual co-host were asked if ``the virtual co-host was distracting'' on a 5-point Likert scale. For the 17 under-participators who were asked questions by the virtual co-host, only 4 participants ``Agreed'' (3 participants) or ''Strongly Agreed'' (1) with that statement. Conversely, 10 said they ``Strongly Disagreed'' (5)  and ``Disagreed'' (5) with the co-host being distracting (see Figure~\ref{fig:distraction}). This suggests that most did not find the virtual co-host distracting, but the variation supported our decision to make the virtual co-host's questions unobtrusive.

\distraction

\hfill\\\noindent\textit{\textbf{Asking mitigated concerns about accuracy.}}

Almost all interviewees said they preferred being asked before interventions. We asked under-participators their opinions on having the virtual co-host ask before intervening. U8 responded that \textit{``I would still prefer [the co-host] to ask questions... sometimes expressions just aren't communicating what you're actually feeling.''} U2 similarly worried that if the system intervened without asking first, it \textit{``might be wrong, and then it negatively impacts the person''} on behalf of whom it is intervening. Participants worried that the AI might inaccurately infer meeting issues where there are none and intervene inappropriately. Asking mitigated this concern, confirming our design intuitions and aligning with other human-AI studies~\cite{kawakami_improving_2022, smith_keeping_2020, binns_human_2022}.

\hfill\\\noindent\textit{\textbf{Asking let participants retain agency.}}

We wondered if asking might have other advantages, so we asked participants to \textbf{assume the virtual co-host could perfectly assess their internal states.} Participants \textit{still} preferred for the virtual co-host to solicit explicit meeting assessments. U8 said \textit{``I like retaining agency, so I would still prefer it to ask questions than try to analyze my face.''} U9 echoed this, saying \textit{``there's that part of autonomy that needs to be maintained, so I wouldn't like the virtual co-host to notify the host without checking in with me first.''} We conclude that OAI is useful for designing AI-based meeting interventions because it improves interventions based on imprecise proxy signals \textit{and} preserves human preferences for agency over AI systems (consistent with ~\citet{schmidt_interactive_2020}). We return to this in the Discussion section.

\hfill\\\noindent\textbf{\textit{\underline{Effectiveness:}}}

\hfill\\\noindent\textit{\textbf{Participants gave the virtual co-host accurate information.}}

Participants reported being honest in their answers to the virtual co-host. U10 even remarked that \textit{``because it's an inanimate being, I'd probably be \textbf{more likely} to give it a very honest answer.''}  This is consistent with prior work showing that people are more likely to disclose difficult information to virtual humans than real humans~\cite{lucas_its_2014}. It also corroborates our ideation participants' intuitions that AI agents can receive critical feedback with minimal discomfort to users, which may make them more effective than humans at gathering information about meeting issues.

\hfill\\\noindent\textit{\textbf{There was a drop-off in engagement with the co-host's questions.}}

Recall that participants reported minimal distraction from the virtual co-host's questions. However, we saw a drop-off in response rate as more feedback was requested. Of the 17 under-participators who received questions from the virtual co-host, 14 answered at least one of the questions, 11 answered the first two, and only 7 answered all three. Further, the virtual co-host's third question was free-form - of the 7 that answered it, only 1 provided unstructured feedback to the host. The 6 other participants just said ``no'' to provide no feedback. On the one hand, this was surprising because multiple ideation session participants said they wanted to provide anonymous feedback to the host. However, participants may not be willing to take on more cognitive load during a meeting to avoid distraction.

\subsubsection{\textbf{\underline{\large Intervene \normalsize}}}\hfill\\

\noindent\textbf{\textit{\underline{Acceptability:}}}

\hfill\\\noindent\textit{\textbf{Participants appreciated socially comfortable interventions.}}

As expected from ideation sessions, participants---both over- and under-participators---preferred that visualizations remain private because public visualizations would create social discomfort. O1 worried that visualizations \textit{``could lead to feelings of embarrassment if it's broadcasted to everybody.''} This may seem self-interested coming from an over-participator, but under-participators shared this concern:

\begin{quote}
\textit{``That puts people on the spot of like, shoot, I've been talking too much or too little. If it's just to the host, they can make space for the person without calling them out.''} (U6)
\end{quote}

U2 similarly said: \textit{``when you give it to everyone, everyone will just speak less and then no one is talking.''} Even when frustrated with an uninclusive meeting, U5 worried that \textit{``if it's a shared display, it might make some folks feel uneasy.''}

\hfill\\\noindent\textit{\textbf{The host has a special mandate.}}

Recall that we selected the host for each meeting at random. Despite this, participants were comfortable with the host having privileged information about everyone's exact speaking times:

\begin{quote}
\textit{Giving it to the host makes it feel like it's a tool for the host to care for everyone, whereas when it's presented to everyone it feels like a monitoring device of your performance.} (H3)
\end{quote}

Non-hosts echoed this. U2 reasoned that the host has \textit{``a mandate to make sure everyone has the room to speak.''} O1 expressed that the host should receive private feedback and visualizations \textit{``because they have responsibilities over other people.''} Importantly, U3 cautioned that \textit{``if you have a boss who doesn't like you, they can use it against you,''} indicating that users' comfort might depend on the meeting host demonstrating positive intentions.

\subjectiveratings

\hfill\\\noindent\textbf{\textit{\underline{Effectiveness:}}}

\hfill\\\noindent\textit{\textbf{Groups with the virtual co-host rated meetings better.}}

Participants in treatment groups rated meetings better overall (Table~\ref{tab:subjectiveratings}). While we would expect this to mean meeting outcomes were improved by the virtual co-host's interventions, our findings on effectiveness paint a different picture.

\hfill\\\noindent\textit{\textbf{Participants paid less attention to interventions than to questions.}}

Even though the virtual co-host's interventions did not require active interaction (unlike the co-host's questions), participants who received interventions varied much more in their desire to redirect attention to its messages. H5 said \textit{``it was just another thing to look at when there were a lot of things going on.''} H4 did not read the messages \textit{``to make sure I was staying focused''}, but did look at the visualizations and could recall the information in them. The other two hosts, H1 and H2, reported paying close attention to the information in the virtual co-host's messages. H2, for example, \textit{``skimmed through the suggestions''}, which served as a \textit{``quick reminder that I need to actively pull people into this conversation''}. H1 read the co-host's instructions for how to stop its messages, but \textit{``chose not to stop it just because I still think they're helpful.''}

Only 1 of 3 over-participators reported looking at the virtual co-host's messages, compared to 4 of 4 hosts. Despite getting reminders from the virtual co-host every 4 minutes, 2 of 3 over-participators missed all its messages. This is likely because they were singularly focused on the meeting goals, or perhaps, not focused on the overall tenor of the meeting and goals of inclusion. This suggests that over-participator interventions may need to be more prominent, harder to ignore, or tailored to their contexts. 

\hfill\\\noindent\textit{\textbf{Participants rationalized away the virtual co-host's feedback.}}

Despite finding co-host messages helpful, hosts did not feel compelled to act on the recommendations. H1 explained that the virtual co-host emphasized the value of running meetings inclusively, but that they were already being inclusive anyway:

\begin{quote}
\textit{It wasn't a huge change for me. When you're leading a meeting, one of the most important parts is getting everybody to speak, so I've always emphasized that.}
\end{quote}

In this case, H1's positive evaluation of the meeting may have been accurate; in their meeting, U4 was the largest under-participator and did express that they thought the meeting was going well, saying \textit{``I did not need to add my voice''} in our follow-up interview. However, it is not clear that H1 could intuitively know U4's meeting experience, given that U4 \textbf{had only spoken for 10 seconds} at the time of intervention (over 8 minutes into the meeting). That H1 maintained a positive self-assessment despite the virtual co-host's assertion that ``some people have not felt comfortable expressing themselves'' is consistent with what we observed across participants.

During another session, the facilitating researcher noticed that H4 was moving quickly through the activity items without allowing time for others to interject---H4's session was shortest by far (approximately 13 minutes, compared to the median meeting length of 28 minutes). H4 also spoke the largest percentage of the meeting compared to other hosts (68.7\%). When asked about the value of including an under-participator in the meeting as the virtual co-host recommended, H4 responded that \textit{``I feel like, in that scenario, if [person] had something to say, [they] would have said it.''} Indeed, H4 argued that they were doing a good job consulting other meeting participants:

\begin{quote}
\textit{``Every time I was trying to make a decision in the meeting I would ask people what they thought. I didn't feel that anybody was hogging the talking time.''}
\end{quote}

The one over-participator who did see the virtual co-host's messages exemplifies how outside perceptions of under-participators' meeting experiences are often inaccurate in self-serving ways. O1 rationalized away the virtual co-host's feedback:

\begin{quote}
\textit{``So when I got the messages saying I might be speaking a little bit more, I was like... I feel like I'm being fair to people and asking for input. I'm talking, but I'm also conceding some of my opinions to other people.''}
\end{quote}

Meanwhile, U5 expressed frustration with O1:

\begin{quote}
\textit{``People were being very rapid-fire, and they weren't leaving a lot of time for discussion. It just kind of seemed like, is everyone fine with this?''}
\end{quote}

Despite reminders from the virtual co-host indicating problems in the meeting, participants mostly ignored its constructive feedback and suggestions. This can be explained by the psychological literature on cognitive dissonance~\cite{harmon-jones_introduction_2019}, which shows that humans may re-orient their beliefs to maintain psychological comfort. That participants so consistently engaged in cognitive dissonance here was somewhat surprising given ideation session participants' intuitions and prior research where awareness feedback successfully changed over-participators' behaviors~\cite{dimicco_impact_2007}. We return to this in the Discussion section for additional analysis.

Low social pressure from AI is the most plausible cause of our interventions' failures, consistent with recent work by~\citet{vollmer_children_2018} finding that robots could not subtly peer pressure adults as effectively as other adults could. Recall that some of our interventions were artificially triggered by researchers. Could hosts and over-participators have simply ignored the feedback because they (correctly) intuited its disconnect from real meeting conditions? The example of O1 complicates this interpretation; O1 rationalized away feedback even when it was informed by U5's deep frustration. Such lack of insight into others' experiences is consistent with recent work by~\citet{houtti_all_2023}, who note that meeting leaders are often completely unaware when their meeting facilitation decisions make others feel excluded. Further, it is difficult to imagine the same feedback from a \textit{human} peer failing to override O1's mistaken assumption in this case---or even \textit{accurate} assumptions in other participants' cases. Indeed, the famous Asch conformity experiments~\cite{asch1956studies} showed that people can even be influenced to give an obviously incorrect answer to a question if it conforms to others' answers.

In light of the virtual co-host's ineffectiveness, our finding that it improved meeting assessments suggests that its mere presence, and perhaps its questions\footnote{Our quantitative results do not meaningfully change when we compare control groups against treatment groups where the co-host was present but did \textit{not} intervene with visualizations and feedback. (See Table~\ref{tab:suppnointervene}.) Alongside the fact that most participants did not receive interventions in the first place, this supports our assertion that the co-host's presence or questions, not its interventions, improved meeting perceptions.}, primed participants to rate meetings better, even if no gains were seen in meeting quality. Our conclusion here follows from recent work noting the presence of a robust AI ``placebo effect''~\cite{kosch_placebo_2023}. This highlights an important issue, which we also return to in the Discussion: deploying AI systems can make participants \textit{feel} better about meetings even if they have no meaningful effects on meeting behaviors.

\subsection{Summary}

Most of the key premises underlying our design insights were borne out in the evaluation. AI systems can (and the virtual co-host did) proactively solicit feedback during meetings. Gentle and private interventions showed more mixed results; the specific interventions we employed were acceptable for the reasons we anticipated (maintaining anonymity and social comfort), but did not improve participants' meeting behaviors. Our qualitative findings shed light on underlying reasons, which we now discuss alongside other takeaways.

\section{Discussion}

Our findings demonstrate the OAI framework's potential in helping us design acceptable and effective AI systems for meeting facilitation. We highlight contributions of the OAI framework as a whole and takeaways from the mixed results of our specific intervention strategy. We synthesize our findings alongside prior work to suggest ways in which future interventions could be more effective, while retaining the features of our specific interventions that made them acceptable to participants.

While our ideation sessions focused on a system to make VC meetings more inclusive, we believe our findings can provide more general guidance. Therefore, the first two Discussion subsections consider the general design implications of findings on the Ask and Intervene steps in OAI. (We do not focus on Observe given the extensive and high-quality prior work on AI observation and inference---e.g. ~\cite{samrose_meetingcoach_2021,murali_affectivespotlight_2021,langner_eyemeet_2022,hu_dark_2023}.) We then highlight implications for design to mitigate inequities in meetings.

\subsection{Ask}

\subsubsection{Asking makes AI inferences more acceptable \textbf{and} more accurate}

Designing the virtual co-host to ask questions mid-meeting conflicted with our intuition to minimize distractions. Indeed,~\citet{liao_what_2016} found that users dislike unsolicited proactive interaction from an automated agent, especially if they have busy work schedules. However, we reasoned this aversion may not apply in the VC context where users already expect synchronous engagement with others verbally and through text-based chat. Enacting this user preference in our implementation let us test whether similar barriers would materialize in such contexts.

Our results suggest proactive ``asking'' during synchronous group work actually makes AI systems \textit{more} acceptable. When asked to choose between a virtual co-host that asks questions and one that simply intervenes autonomously, all but one under-participator preferred the former---even after having just experienced a meeting in which they had the burden of answering its questions. Preferences were driven by improved sense of agency and reduced concerns about accuracy.

Accuracy was a major concern for users in prior work, and asking provides a means of overcoming it. A user's subjective impression of their inclusion in a meeting is more reliable than what an algorithm could produce by analyzing their face or speech. This could be used to augment AI inferences. For example, MeetingCoach's post-meeting dashboard shows a breakdown of a user's tone throughout the meeting~\cite{samrose_meetingcoach_2021}. What if MeetingCoach were designed to ask assessment questions when a speaker had a negative tone? This would enable the system to confirm AI inferences before acting on them, overcoming a key limitation of AI systems \textit{and} making it more acceptable. Explicit responses could also be displayed to users \textit{alongside} inferential data to substantiate inferences.

\subsubsection{Design guidelines for asking}

Our results suggest several design guidelines for systems that ask questions during meetings:

\begin{itemize}
\item Systems should distribute the burden of answering across multiple users, rather than just a single one. When possible, interventions should not be contingent on getting answers to \textit{all} questions.
\item As expected, it was easier to get the attention of under-participators than over-participators through text-based chat. \textbf{Systems that target middle- or over-participators will have to do more work to make questions visually prominent and difficult to miss.}
\item Despite ideation session participants wanting the ability to give free-form feedback, participants in our meetings rarely gave feedback mid-meeting. \textbf{Questions should be yes-or-no, to minimize cognitive burden.} We hypothesize that Likert scale questions can be considered if the granularity justifies the likely reduction in response rate.
\end{itemize}

\subsection{Intervene}

\subsubsection{Effective interventions should make social pressure more salient}

By observing the effects of speaking time visualizations in meetings, ~\citet{dimicco_impact_2007} concluded that group awareness feedback will cause individuals to self-regulate and conform with the feedback's \textit{implied} norms. Our virtual co-host made normative expectations more \textit{explicit}; it directly messaged over-participators and hosts to indicate when they were not meeting normative expectations. Yet its constructive criticism and suggestions fell mostly on deaf ears. Importantly, DiMicco et al.'s meeting visualizations differed in that they \textit{were shared among all meeting participants}.

We conclude that awareness feedback in meetings likely achieves most of its effects by augmenting perceived normative pressure. Individuals inherently understand that over-participating can lead others to judge them negatively. Making the \textit{judgment} more salient (by having a \textit{public} speaking display) is more effective than making the \textit{over-participation} more salient (in our case, by having an AI agent provide \textit{private} feedback and visualizations). We speculate that highlighting normative pressure from others reduces the risk of cognitive dissonance~\cite{harmon-jones_introduction_2019} because it does not require users to accept that critical information about themselves is accurate---only that others may perceive it as accurate. Awareness feedback interventions that seek to influence meeting participants' behavior should therefore be designed to emphasize the possibility of judgment from peers. 

There is a fundamental tension here: users prefer systems that \textit{minimize} social pressure, but social pressure is needed to prompt behavior change. This tension is at the heart of why our interventions failed to change participants' behavior. However, future systems can be creative in balancing these concerns. For example, we could take a phased approach that relies on the \textit{threat} of public interventions---e.g., where participants are told that private interventions will become public \textit{only} if not acted on. This would maintain social comfort, while incentivizing users to change their behavior to \textit{continue} maintaining social comfort. Alternatively, making participants aware that the co-host is acting on behalf of another meeting participant could reduce the risk of cognitive dissonance. AI feedback would be understood to reflect a real human's assessment, prompting behavior change, while still minimizing social pressure because the feedback is delivered indirectly. These interventions have potential in theory, and should be empirically tested in future work.

\subsubsection{Computers are social actors with less social pressure}

Our findings build further on Nass et al.'s seminal work demonstrating that humans subconsciously interact with machines as if they were social beings~\cite{nass_computers_1994}. Our results suggest that AI agents cannot apply normative pressure to the same degree as humans situated in a similar social role, making it more difficult for them to overcome psychological biases like cognitive dissonance~\cite{harmon-jones_introduction_2019}.

In cases where normative pressure is needed, we should design AI systems to communicate desired norms more aggressively. For example, our virtual co-host could have been more effective if it prompted participants to explicitly acknowledge the norm violations---e.g., by preventing them from un-muting until they acknowledged the feedback and their intention to act on it. Alternatively, AI systems could rely on other humans to give their interventions normative weight. Our results suggest, for example, that an AI fitness accountability coach might not be as effective as a human coach even if it competently performed all the same functions. This system could rely on normative pressure from other humans by, e.g., sharing the user's progress with their friends (just as many modern fitness apps do).

While low normative pressure is a \textit{limitation} of AI agents in the intervention use-case, it also could be an \textit{advantage} when we \textit{want to} minimize normative pressure. Soliciting meeting assessments is one case---recall that one of our participants said it was easier to give the virtual co-host negative feedback \textbf{because it was virtual}. This strength can be leveraged in other domains where giving feedback is difficult. For example, Google's Q\&A tool for all-hands meetings, which combines and summarizes employee questions, reportedly doubled the number of employees contributing feedback~\cite{quiroz-gutierrez_google_2024}.\footnote{We should note that the tool has been criticized for ``softening'' questions to leadership. As evidenced by our findings, however, friendliness can be valued over directness in other organizational contexts.} Future work could explore other applications of AI as a means of facilitating the exchange of feedback in groups.

\subsubsection{Positive priming can make ineffective interventions seem effective}

The virtual co-host's positive priming effects demonstrate the imperative to measure systems' effects on meetings based on tangible outcomes, self-reported subjective data, \textit{and} in-depth qualitative inquiry. In its current form, we expect our system would marginally improve real workers' subjective meeting experiences despite not truly improving meeting behaviors in any meaningful way.

Researchers and practitioners should be mindful of priming when users are involved in the design process---e.g., through value-sensitive design~\cite{friedman_value_2001}, user-centered design~\cite{abras_user-centered_2004}, participatory design~\cite{schuler_participatory_1993}, or other methods---especially when evaluating systems that seek to challenge users or alter their behavior. Recall that following participants' preferences to intervene privately made our system both \textit{more} acceptable and \textit{less} effective. It is not entirely surprising that participants designed a system they liked, and that did not sufficiently constrain their negative behaviors. We hypothesize that systems designed with user involvement are more likely to see positive subjective outcomes alongside negative/neutral objective outcomes, though more work is needed to confirm or reject this hypothesis. 

Importantly, it may not be sufficient to consult more (or more diverse) stakeholders. Recall that \textit{even under-participators} preferred interventions to minimize \textit{others'} social discomfort. This may have been caused by vicarious embarrassment~\cite{krach_your_2011}---i.e., under-participators may feel uncomfortable with someone else being ``called out'' in their presence. Said another way, consulting the stakeholders who stood to benefit the most did not suggest different design conclusions. While this may be a limitation of user-centered methods (and should be investigated more), it should not deter us from involving users in the design process. Our systems should be both effective \textit{and} acceptable, so subjective outcomes are still important to assess.

\subsection{Implications for Future Work on VC Meeting Inclusion}

So far, our discussion has centered around the significance of our findings for designing AI-based meeting systems in general. We believe this will be valuable to researchers in CHI; however, our findings also include important takeaways about the specific problem of inclusion in meetings. We conclude by connecting our implications to this context. We offer these as methodological and design guidelines for future work on designing AI systems to make meetings more inclusive.

\subsubsection{High risk of cognitive dissonance}

Recall that participants engaged in cognitive dissonance because our system could not apply sufficient social pressure. Future researchers should anticipate and design for this risk. Prior work in psychology extensively documents the self-serving bias---\textit{``a tendency for people to take personal responsibility for their desirable outcomes yet externalize responsibility for their undesirable outcomes.''}~\cite{shepperd_exploring_2008} Similarly, we expect that users will tend to externalize responsibility for adverse meeting outcomes, increasing the risk of cognitive dissonance when we seek to change behavior by surfacing their roles in making the meeting less inclusive.

\subsubsection{Surreptitiously perpetuating inequity}

Our findings demonstrate the imperative for researchers and designers to evaluate systems' effects holistically. In its current form, we imagine our virtual co-host would marginally improve real workers' subjective perceptions of meeting inclusion and, in doing so, \textbf{unintentionally exacerbate inequities} by making people complacent. Why worry about meeting inclusion when we have an AI tool that mitigates it? (Except, of course, it did not.) Given how contextual, multi-faceted, and difficult-to-measure inclusion is, these conflicting effects can be difficult to identify unless we are looking for them. Thus, narrow evaluations risk leading us to design systems that surreptitiously perpetuate the very harms we intend to mitigate.

\section{Limitations}

Our chosen methods, while appropriate for the questions we sought to investigate, have several inherent limitations. 

\subsection{Controlled Environment}

A lab study enabled us to compare the experiences of groups in a somewhat controlled environment, but real meetings are far from controlled and are therefore impossible to emulate perfectly. For example, we could expect that meeting leaders who feel a strong sense of responsibility over their team meetings might be more receptive to a virtual co-host's negative feedback (and therefore less likely to rationalize its feedback away). 

Further, a lab study precluded us from examining use over time. In organizational settings, we might see hosts rationalize away feedback in the meeting itself, but then seek out feedback from under-participators after the meeting. This might in turn affect how they react to subsequent feedback from the virtual co-host. These limitations suggest avenues for further research, ideally in the form of field studies where AI-based systems that employ OAI are tested in real meetings over time.

\subsection{Between-Subjects and Inter-Group Differences}

Due to our between-subjects design, differences in meeting assessments could simply reflect inherent differences between groups, rather than effects of the treatment. While random assignment accounts for this issue in theory, recall that we assigned participants \textit{pseudo}-randomly---to ensure balance between conditions, then to create additional treatment groups with researcher-triggered interventions. While we see no obvious reasons this group assignment strategy could have systematically biased group composition, this possibility cannot be ruled out entirely.

\subsection{Interactions Between OAI Phases}

Interactions between phases of OAI were beyond the scope of our study. When designing an OAI-based system, we should be mindful of how implementation decisions in one phase could affect reactions to an entirely different phase. For example, if Ask were timed to overtly follow specific behavioral cues, this could make Observe more salient to users such that they do experience the ``watching-eye''~\cite{hu_dark_2023} effect contra our findings. And as already mentioned, the genuineness of responses in Ask could change how people react to Intervene, especially if they intuit a disconnect between the AI's interventions and real meeting conditions. Ideally, these and other interactions could be tested via 2x2 factorial studies---but at the very least, they should be considered when designing other OAI-based systems.

\subsection{Other Cultural Contexts}

Our participants were all US-based English-speakers. Many of our findings may not transfer to other cultural contexts. For example, the ``Ask'' phase could plausibly fail to get honest responses in cultures with high power distance~\cite{hofstede2011dimensionalizing}. Future work should consider this possibility when employing or studying OAI-based systems in cultural contexts that are substantially different from the one we focused on here.

\section{Conclusion}

Through user-centered ideation sessions and an exploratory lab study, this work contributes details and operationalization of a new framework, ``Observe, Ask, Intervene'' (OAI), for designing AI tools to mediate synchronous group interactions. Our work provides guidelines for designing proactive AI agents to support group work, especially in the context of improving meeting inclusion.

\clearpage

\bibliographystyle{ACM-Reference-Format}
\bibliography{sample-base}

\clearpage
\appendix
\beginsupplement
\section{Appendix}

\begin{table}[htbp]
\centering
\def\arraystretch{1.5}
\begin{tabular}{|l|r|r|r|r|r|r|r|}
\hline
\textbf{Construct} & \textbf{\textit{M} (control)} & \textbf{\textit{SD} (control)} & \textbf{\textit{M} (treatment)} & \textbf{\textit{SD} (treatment)} & \textit{\textbf{df}} & \textit{\textbf{t}} & \textit{\textbf{p}} \\ \hline
Communication & 8.45 & 3.37 &  \textbf{6.18} & 2.96 & 55.90 & 2.89 & **0.005 \\ \hline
Discussion Quality & \textbf{17.24} & 2.18 & 17.10 & 2.62 & 65.11 & 0.24 & 0.81 \\ \hline
Status Effects & 7.83 & 3.75 & \textbf{6.05} & 2.84 & 50.19 & 2.14 & *0.04 \\ \hline
Teamwork & 12.38 & 1.86 & \textbf{12.90} & 1.79 & 59.15 & -1.15 & 0.25 \\ \hline
Efficiency & 7.69 & 1.39 & \textbf{8.92} & 1.26 & 57.09 & -3.76 & ***0.0004 \\ \hline
\end{tabular}
\vspace{10pt}
\caption{Same as Table~\ref{tab:subjectiveratings}, but with the modified Efficiency construct (using only questions E2 and E3) included. Welch's two sample $t$-tests of user assessments on the five meeting quality constructs in Davison~\cite{davison_instrument_1999}. Participants who had the co-host present rated meetings better on all but one construct. 5-point Likert scale responses were summed within each construct to derive a corresponding value for each participant. Note that some of the constructs are worded positively while others are worded negatively~\cite[see~A.1]{davison_instrument_1999}, so the ``better'' mean value for each row is indicated in bold.}
\Description{Table S1: Contains t-test results for five meeting constructs. From left to right, the columns represent the construct, the mean response for control groups, the standard deviation for control groups, the mean response for treatment groups, the standard deviation for treatment groups, the degrees of freedom, the t-value, and the p-value of a two-sample Welsh's $t$-test.}
\label{tab:suppmain}
\end{table}

\begin{table}[htbp]
\centering
\def\arraystretch{1.5}
\begin{tabular}{|l|r|r|r|r|r|r|r|}
\hline
\textbf{Construct} & \textbf{\textit{M} (control)} & \textbf{\textit{SD} (control)} & \textbf{\textit{M} (treatment)} & \textbf{\textit{SD} (treatment)} & \textit{\textbf{df}} & \textit{\textbf{t}} & \textit{\textbf{p}} \\ \hline
Communication & 8.45 & 3.37 &  \textbf{5.71} & 2.52 & 41.22 & 3.14 & **0.003 \\ \hline
Discussion Quality & \textbf{17.24} & 2.18 & 17.05 & 2.30 & 32.17 & 0.26 & 0.79 \\ \hline
Status Effects & 7.83 & 3.75 & \textbf{5.47} & 1.62 & 41.37 & 2.94 & **0.005 \\ \hline
Teamwork & 12.38 & 1.86 & \textbf{13.35} & 1.58 & 38.18 & -1.89 & 0.07 \\ \hline
Efficiency & 7.69 & 1.39 & \textbf{9.06} & 1.03 & 41.46 & -3.81 & ***0.0005 \\ \hline
\end{tabular}
\vspace{10pt}
\caption{Same as Table~\ref{tab:subjectiveratings}, but with the modified Efficiency construct (using only questions E2 and E3) included \textit{and} excluding treatment groups where the virtual co-host intervened. Welch's two sample $t$-tests of user assessments on the five meeting quality constructs in Davison~\cite{davison_instrument_1999}. Participants who had the co-host present rated meetings better on all but one construct, \textit{even if} we only look at meetings where the co-host did not intervene. 5-point Likert scale responses were summed within each construct to derive a corresponding value for each participant. Note that some of the constructs are worded positively while others are worded negatively~\cite[see~A.1]{davison_instrument_1999}, so the ``better'' mean value for each row is indicated in bold.}
\Description{Table S2: Contains t-test results for five meeting constructs. From left to right, the columns represent the construct, the mean response for control groups, the standard deviation for control groups, the mean response for treatment groups, the standard deviation for treatment groups, the degrees of freedom, the t-value, and the p-value of a two-sample Welsh's $t$-test.}
\label{tab:suppnointervene}
\end{table}

\end{document}